# Lattice dynamics in mono- and few-layer sheets of WS$_2$ and WSe$_2$

*Weijie Zhao*[a,c,#], *Zohreh Ghorannevis*[a,c,#], *Amara Kiran Kumar*[b], *Jing Ren Pang*[b], *Minglin Toh*[d], *Xin Zhang*[e], *Christian Kloc*[d], *Ping Heng Tan*[e], *Goki Eda*[a,b,c,*]

[a] Department of Physics, National University of Singapore, 2 Science Drive 3, Singapore 117542

[b] Department of Chemistry, National University of Singapore, 3 Science Drive 3, Singapore 117543

[c] Graphene Research Centre, National University of Singapore, 6 Science Drive 2, Singapore 117546

[d] School of Materials Science and Engineering, Nanyang Technological University, N4.1 Nanyang Avenue, Singapore 639798

[e] State Key Laboratory of Superlattices and Microstructures, Institute of Semiconductors, Chinese Academy of Sciences, Beijing 100083, China

[*] E-mail: g.eda@nus.edu.sg

**Abstract**

Thickness is one of the fundamental parameters that define the electronic, optical, and thermal properties of two-dimensional (2D) crystals. Phonons in molybdenum disulfide (MoS$_2$) were recently found to exhibit unique thickness dependence due to interplay between short and long range interactions. Here we report Raman spectra of atomically thin sheets of WS$_2$ and WSe$_2$ , isoelectronic compounds of MoS$_2$, in the mono- to few-layer thickness regime. We show that, similar to the case of MoS$_2$, the characteristic $A_{1g}$ and $E_{2g}^1$ modes exhibit stiffening and softening with increasing number of layers, respectively, with a small shift of less than 3 cm$^{-1}$ due to large mass of the atoms. Thickness dependence is also observed in a series of multiphonon bands arising from overtone, combination, and zone edge phonons, whose intensity exhibit significant enhancement in excitonic resonance conditions. Some of these multiphonon peaks are found to be absent only in monolayers. These features provide a unique fingerprint and rapid identification for monolayer flakes.

**Introduction**

Two-dimensional (2D) crystals derived from inorganic layered compounds offer a unique platform to explore fundamental condensed matter phenomena[1]. Recently, tremendous interest has focused on 2D crystals of molybdenum disulfide ($MoS_2$) and other members of layered transition metal dichalcogenides (LTMDs) due to their intriguing electrical and optical properties[2-13]. A single monolayer of $MoS_2$ is a direct gap semiconductor with high in-plane carrier mobility and excellent gate coupling for electrostatic control of charge carrier density[10, 11, 14]. Finite band gap and remarkable device performance make $MoS_2$ a complementary 2D material to graphene in nanoelectronics and photonics[11, 15]. Further, sizable spin-orbit interaction along with broken inversion symmetry in monolayer $MoS_2$ allows optical access to valley degrees of freedom, demonstrating potential in novel spintronic and valleytronic devices[4, 9, 16, 17].

Tungsten-based LTMDs such as $WS_2$ and $WSe_2$ are isoelectronic to $MoS_2$ and exhibit a similar set of intriguing properties in 2D crystal form[7, 11, 18-27]. Their monolayer consists of an X-M-X sandwich[28] (M and X denote transition metal and chalcogen atoms, respectively) with trigonal prismatic coordination as in $MoS_2$. Exfoliation of $WS_2$ and $WSe_2$ to mono- to few-layer thick sheets has been recently demonstrated by many groups[7, 18, 19, 21, 25, 26, 29-33]. This has led to observation of remarkable field-modulated transport with large in-plane mobility[11, 19], indirect-to-direct band gap transition upon isolation of single layers[7, 21, 22, 32], robust valley polarization[25], second harmonic generation[7], and tightly bound trions[25]. Spin-orbit interaction in these materials is substantially larger[34-39] compared to that in $MoS_2$ thus offering a robust platform to study spin and valley physics.

While confinement effects on the electronic and excitonic dispersion relation in atomically thin sheets of tungsten dichalcogenides has been extensively studied to date[21, 34, 39-41]. On the other hand, little has been understood about the phonon behaviors in these 2D crystals. Lee et al.[42] recently reported that phonon frequency of atomically thin MoS$_2$ flakes exhibits unique thickness dependence where two characteristic Raman active modes $A_{1g}$ and $E_{2g}^1$ exhibit opposite trends with thickness. Specifically, the $A_{1g}$ mode, which involves the out-of-plane displacement of S atoms, is found to stiffen with increasing number of layers. In contrast, the $E_{2g}^1$ mode, which involves the in-plane displacement of Mo and S atoms, exhibits softening with flake thickness. Further studies[43] have shown that the shift in the $A_{1g}$ mode can be explained by the interlayer interaction of S atoms in the neighboring planes, while the unexpected trend of the $E_{2g}^1$ mode is explained by dielectric screening of long range Coulomb interaction.

Although these phonon modes in tungsten dichalcogenides were expected to follow the same trends as in MoS$_2$[42, 44, 45], there have been conflicting reports on the thickness dependence of phonon modes in WS$_2$ and WSe$_2$ sheets in mono- to few-layer regime.[7, 22, 26, 31, 32, 46, 47] For WSe$_2$, this is partly attributed to the lack of consensus on the assignment of $A_{1g}$ mode[7, 31, 47]. Of particular interest is the discrepancy between the experimental with the theoretical results where latter suggest that $A_{1g}$ and $E_{2g}^1$ are degenerate in WSe$_2$.[47] Further, behavior of multiphonon modes which contain rich information on the electronic and phonon dispersion relation[46] are largely unexplored. A detailed study of the phonon properties needs to be conducted in order to achieve valuable insight into phonon confinement effects in these intriguing 2D materials.

Here, we report detailed studies on the Raman spectra of mechanically exfoliated mono- and few-layer WS$_2$ and WSe$_2$ flakes. With the use of polarized spectroscopy, we demonstrate that for both materials the $A_{1g}$ and $E_{2g}^1$ modes exhibit opposite shift with increasing number of layers indicating the effects of both short and long range interactions. We find that these phonon modes are degenerate in monolayer WSe$_2$ as predicted by theory[47] but the degeneracy is lifted in multilayers. We further discuss excitonic resonance Raman spectra for WS$_2$ and WSe$_2$ where a series of multiphonon bands are observed. We demonstrate that some of these features contain unique fingerprints of monolayer flakes.

**Results and Discussion**

Figure 1a shows the optical contrast or differential reflectance spectra of 3L WS$_2$ and WSe$_2$ flakes. The optical contrast is defined as ($R_{S+Q}$ - $R_Q$)/$R_Q$ where $R_{S+Q}$ and $R_Q$ are the reflected light intensities from the quartz substrate with and without samples, respectively[2, 48]. This quantity is proportional to absorbance for ultrathin samples and its spectral response can be interpreted as absorption spectrum[2, 21, 48]. Characteristic excitonic absorption peaks A and B are observed along with a higher energy density of states peak C and split exciton peaks A' and B'. The A and B excitonic absorption peaks arise from optical transitions involving spin-orbit split valence band and degenerate conduction band at the K point of the Brillouin zone[34, 49]. In this study, we investigate Raman scattering with three excitation wavelengths (472, 532, and 633 nm). The 532 nm excitation is in resonance with the B exciton peak of WS$_2$ and A' exciton peak of WSe$_2$. On the other hand, 473 nm excitation is roughly in resonance with the interband transition peak C for WS$_2$ and interband absorption continuum for WSe$_2$. The 633 nm excitation is in resonance with the A excitonic absorption for WS$_2$ and interband absorption but close to the B excitonic absorption for WSe$_2$. Raman features of both WS$_2$ and WSe$_2$ strongly depend on the excitation conditions due to energy dependent Raman cross section of the phonons (See Supporting Information for details). The absorption peaks shift slightly with flake thickness[21] but the resonance conditions remain largely unaltered.

The crystal structure of 2H-WX$_2$ belongs to $D_{6h}^4$ point group. There are 18 lattice dynamical modes at the center of the Brillouin zone ($\Gamma$ point)[50-54]. The irreducible representations of zone center phonons are as follows[50, 51, 54]:

$$\Gamma = A_{1g} + 2A_{2u} + B_{1u} + 2B_{2g} + E_{1g} + 2E_{1u} + E_{2u} + 2E_{2g}$$

The atomic displacement of the four Raman active modes $A_{1g}$, $E_{1g}$, $E_{2g}^1$ and $E_{2g}^2$ is shown in Figure 1b. The $A_{1g}$ mode is an out-of-plane vibration involving only the chalcogen atoms while the $E_{2g}^1$ mode involves in-plane displacement of transition metal and chalcogen atoms[50, 51]. The $E_{2g}^2$ mode is shear mode corresponding to the vibration of two rigid layers against each other and appears at very low frequencies (< 50 cm$^{-1}$)[26, 45, 50, 51, 53, 55-58]. The $E_{1g}$ mode, which is an in-plane vibration of only the chalcogen atoms, is forbidden in the back-scattering Raman configuration[50, 51].

Monolayer WX$_2$ belongs to $D_{3h}$ point group and has 9 modes at the Brillouin zone center.[43] The rigid layer shear mode $E_{2g}^2$ is absent in monolayers[26, 45, 55, 58]. Unpolarized Raman spectrum of bulk WS$_2$ obtained with 473 nm excitation shows characteristic $A_{1g}$ and $E_{2g}^1$ peaks that are clearly separated and of similar intensity (Figure 1c). In contrast, only one prominent peak can be clearly seen in bulk WSe$_2$ spectrum in the frequency region where we expect $A_{1g}$ and $E_{2g}^1$ peaks[57] (See Supporting Information). In 633 nm excitation condition, however, multiple peaks are evident in this region (Figure 1d). There have been inconsistent reports on the assignment of these peaks[7, 31, 32, 47]. We demonstrate below that two peaks found at 248 and 250 cm$^{-1}$ are $E_{2g}^1$ and $A_{1g}$ peaks in agreement with the previous study by Mead and Irwin[57].

In polarized back-scattering Raman spectroscopy, $A_{1g}$ mode is allowed in parallel polarization ($Z(XX)\bar{Z}$) but forbidden in cross polarization ($Z(XY)\bar{Z}$) conditions[45]. Thus $A_{1g}$ mode can be identified by observing Raman spectra in the two polarization conditions. Figure 2a shows the parallel and cross polarized Raman spectra of 1 to 5L and bulk WS$_2$ flakes obtained with 473 nm excitation. The disappearance of the peak at ~ 420 cm$^{-1}$ in cross polarization confirms that it is a $A_{1g}$ mode. The peak at around ~ 356 cm$^{-1}$ with no polarization dependence is a $E_{2g}^1$ mode. The peak positions agree well with the previous reports[53, 54]. As shown in Figure 2, the $A_{1g}$ phonon stiffens and $E_{2g}^1$ phonon softens with increasing flake thickness similar to the case of MoS$_2$[42] and as predicted by recent theoretical studies[43]. The difference in the frequency of these peaks is 60 and 65 cm$^{-1}$ for monolayer and bulk samples, respectively (Figure 2b). The intensity ratios ($A_{1g} / E_{2g}^1$) show a similar trend as seen in MoS$_2$. The FWHM is found to consistently decrease with increasing number of layers (Figure 2c).

Raman spectra of WSe$_2$ flakes measured in two polarization configurations and with 633 nm excitation are shown in Figure 3a. In contrast to the case in WS$_2$, the parallel polarization spectra of WSe$_2$ flakes exhibit one prominent peak with a small shoulder in the frequency region where $A_{1g}$ and $E_{2g}^1$ peaks are expected to be observed[57]. The main peak upshifts from 249.5 to 251 cm$^{-1}$ with increasing flake thickness from monolayer to bulk. In cross polarized spectra, a single peak is observed in this frequency region. This peak is found to downshift with increasing flake thickness. The peak intensity is normalized in Figure 3a to show this trend. The opposite thickness dependence and polarization dependence indicates that they are indeed $A_{1g}$ and $E_{2g}^1$ modes. The peak at 257 cm$^{-1}$ did not show consistent polarization dependence, as shown in the Figure S2 of the Supporting Information. This peak was reported to be the $A_{1g}$ phonon by some groups[7, 31, 47]. As shown in Figure1 and Figure S2, this peak is very broad, which is uncharacteristic for a first order Raman peak, and shows only weak dependence on the polarization configuration. It is assigned to the 2LA(M) phonon here according the theoritical calculation[59]. Similar to MoS$_2$ and WS$_2$, the $A_{1g}$ ($E_{2g}^1$) phonon of WSe$_2$ stiffens (softens) with increase in flake thickness (Figure 3b). The key difference is that the two modes become virtually degenerate in the single layer limit. We note that in 473 nm excitation condition, only a single peak appears in the unpolarized and parallel polarized spectra due to relatively weak $E_{2g}^1$ peak. Similar to MoS$_2$ and WS$_2$, The general trends in FWHM and intensity ratio of $A_{1g}$ and $E_{2g}^1$ peaks were consistent with those of MoS$_2$ and WS$_2$ (Figure 3c).

Figure 4 shows the comparison of frequencies and shifts for $A_{1g}$ and $E_{2g}^1$ peaks in different Group 6 TMDs. The $E_{2g}^1$ frequency consistently decreases with increasing "molecular weight" of the unit cell. On the other hand, a distinct jump is found in the $A_{1g}$ frequency between the MS$_2$ and MSe$_2$ groups. This reflects the difference in the chalcogen atomic masses. Difference in phonon frequencies between monolayer and bulk flakes ($\Delta A_{1g}$ and $\Delta E_{2g}^1$) is also highly dependent on the material (Figure 4b). Stiffening and softening of the modes are most prominently observed in MoS$_2$ where $A_{1g}$ and $E_{2g}^1$ peaks shift by 4 and 2 cm$^{-1}$, respectively.[42] The shift becomes less pronounced with "heavier" MX$_2$ with one exception of $E_{2g}^1$ mode of MoSe$_2$ which has been reported to exhibit comparatively large shift (~ 3.6 cm$^{-1}$)[18]. Due to the heavy atoms, WSe$_2$ shows the smallest thickness dependent shift (~ 1.3 cm$^{-1}$) for the two modes.

The small thickness dependence of the $A_{1g}$ and $E_{2g}^1$ peaks in WS$_2$ and WSe$_2$ imply that they are not ideal fingerprints for identifying the number of layers. In the following, we discuss thickness dependence of other phonons that are observed in excitonic resonance conditions. We focus the following discussions on 532 nm resonance. Details of results obtained with 633 nm excitation are presented in the Supporting Information.

Figures 5 and 6 show unpolarized Raman spectra obtained with 532 nm excitation, which is in resonance with B and A' exciton absorption peaks for WS$_2$ and WSe$_2$, respectively. A series of overtone and combination peaks arising from the Brillouin zone center and zone edge phonons are observed along with the first order $E_{2g}^1$ and $A_{1g}$ modes. Detailed assignments of multiphonon bands according to recently calculated phonon dispersion curves[43, 59] are summarized in Table S1 of the Supporting Information.

The most prominent resonance feature emerges near the $E_{2g}^1$ peak for both WS$_2$ and WSe$_2$. This peak, labeled as 2LA(M), is a second-order Raman mode due to LA phonons at the M point in the Brillouin zone[43, 54]. We found that the 2LA(M) mode shows distinct downshift of 5 cm$^{-1}$ with increasing flake thickness from monolayer to trilayer for WSe$_2$ (See Supporting Information for details). In contrast, the same mode in WS$_2$ showed no obvious trend. The first-order phonon LA(M) in both WS$_2$ and WSe$_2$ is widely involved in the overtone and combination mode of other zone edge phonons, similar to the case of MoS$_2$[44, 60].

We observed similar multiphonon bands and 2LA(M) features with 633 nm excitation, which is in resonance with A and B exciton absorption for WS$_2$ and WSe$_2$, respectively (See Supporting Information for details). These results indicate that the enhancement of the total Raman cross section at excitonic resonance in which excitons serve as the intermediate state is stronger compared to that of interband resonance. The strong enhancement at excitonic resonance is attributed to the characteristics of excitons in layered materials such as large binding energy, enhanced oscillator strength, and small damping constant[28, 34, 61-63]. It should be noted that resonance Raman features are also seen with 473 nm excitation but to a lesser degree (See Supporting Information for details).

Figure 5b highlights that a peak at ~ 310 cm$^{-1}$, which is consistently observed for multilayer flakes of WS$_2$, is absent in monolayers. The origin of this mode is not clear, however, its absence in monolayers suggests that it may be related to rigid layer shear. Relatively large frequency of this peak suggests that it is a combination of low-frequency modes[26, 45] with another phonon mode. Figure 6b shows that similar behavior is observed in a peak located around 308 cm$^{-1}$ for WSe$_2$. This peak is absent in monolayers and shows clear softening with increasing thickness. Thus, it is possibly a combination mode of a shear and $E_{2g}^1$ modes. Further evidence is required to verify our speculations on the origin of the peak. These features are useful for quickly identifying WS$_2$ and WSe$_2$ monolayers.

**Conclusions**

In summary, we study Raman scattering in mono- and few-layer WS$_2$ and WSe$_2$ flakes. Characteristic $A_{1g}$ and $E_{2g}^1$ phonon modes show distinct thickness dependence where the former stiffens and the latter softens with increasing number of layers. While the general behaviors are similar to the case of MoS$_2$, the thickness dependent shift of $A_{1g}$ and $E_{2g}^1$ peaks is considerably smaller for WS$_2$ and WSe$_2$ possibly due to the larger atomic masses. The $A_{1g}$ and $E_{2g}^1$ modes in WSe$_2$ exhibit small frequency difference in multilayers and become degenerate in single layers. Presence of $E_{2g}^1$ peak can be verified in cross polarized spectra where the $A_{1g}$ mode is forbidden. Excitonic resonance Raman scattering reveals a series of multiphonon bands involving the Raman inactive phonons at Brillouin zone center and zone edge phonons. We demonstrate that some of the multiphonon bands are absent in monolayers but activated in multilayers, suggesting possible contributions from the low frequency modes. These features allow rapid and unambiguous identification of monolayers.

**Methods**

Bulk crystals of 2H-WS$_2$ and 2H-WSe$_2$ were grown by chemical vapor transport (CVT). Bulk crystals of 2H-WS$_2$ and 2H-WSe$_2$ crystals were mechanically exfoliated on to quartz substrates and used for subsequent spectroscopic characterization. The number of layers in the flakes was confirmed by optical contrast and atomic force microscope (AFM)[21]. The optical contrast was measured using a tungsten-halogen lamp coupled to a Raman spectrometer. Raman spectra were acquired in ambient conditions using 473, 532 and 633 nm laser excitations. The laser power on the sample during Raman measurement was kept below 150 µW in order to avoid sample damage and excessive heating. A 2400 grooves/mm grating was used to achieve spectral resolution of below 1 cm$^{-1}$. The silicon Raman mode at 520 cm$^{-1}$ was used for the calibration prior to measurements.

**Acknowledgment**

G Eda acknowledges Singapore National Research Foundation for funding the research under NRF Research Fellowship (NRF-NRFF2011-02). PH Tan thanks the supports from NSFC under grants 10934007 and 11225421.

**Supporting Information**

Detailed Raman spectra obtained with three laser excitations and possible assignments of the multiphonon bands are presented in the Supporting Information.

Figures and Figure captions

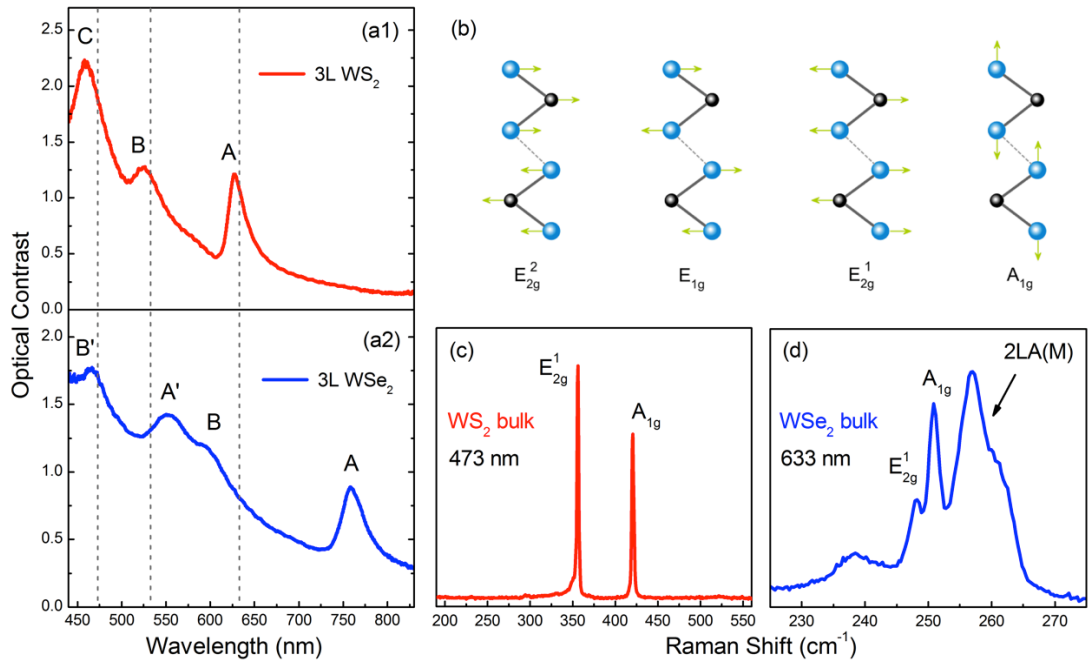

Figure 1 - Optical contrast of 3L (a1) WS$_2$ and (a2) WSe$_2$ flakes. The grey dashed arrows indicate wavelength of the excitation lasers used for Raman measurements. (b) Schematics showing atomic displacement of four Raman active modes in WS$_2$ and WSe$_2$. (c-d) Unpolarized Raman spectra of bulk (c) WS$_2$ and (d) WSe$_2$ obtained with 473 and 633 nm laser excitation, respectively.

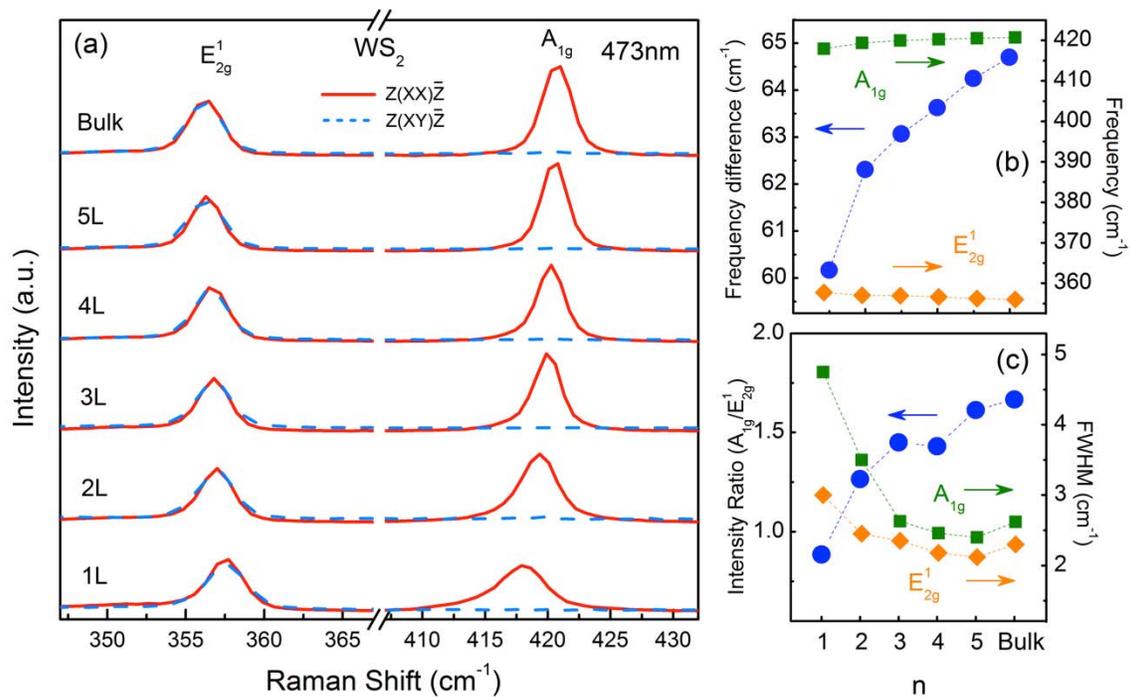

Figure 2 - (a) Raman spectra of 1 to 5L and bulk $WS_2$ flakes obtained in the parallel ($Z(XX)\bar{Z}$) and cross ($Z(XY)\bar{Z}$) polarization conditions obtained with 473 nm excitation. The spectra are normalized and vertically offset for clarity. (b) Position of the $A_{1g}$ and $E_{2g}^1$ modes (right vertical axis) and their difference (left vertical axis) as a function of the number of layers (n). (c) Intensity ratio (left vertical axis) and FWHM (right vertical axis) of $A_{1g}$ and $E_{2g}^1$ modes as a function of the number of layers. The spectral resolution is about 0.8cm$^{-1}$.

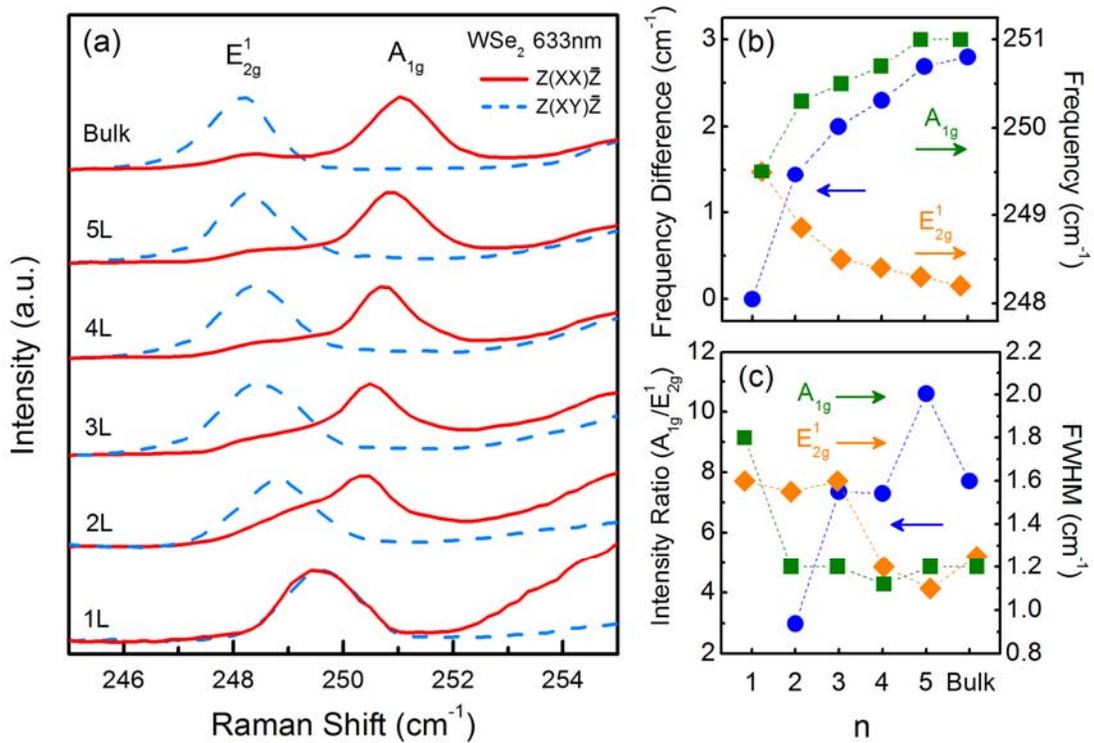

Figure 3 - (a) Raman spectra of 1 to 5L and bulk WSe$_2$ obtained in the parallel ($Z(XX)\bar{Z}$) and cross ($Z(XY)\bar{Z}$) polarization conditions obtained with 633 nm excitation. The spectra are normalized and vertically offset for clarity. (b) Position of the $A_{1g}$ and $E_{2g}^1$ modes (right vertical axis) and their difference (left vertical axis) as a function of the number of layers (n). (c) Intensity ratio (left vertical axis) and FWHM (right vertical axis) of $A_{1g}$ and $E_{2g}^1$ modes as a function of the number of layers. The spectral resolution is about 0.2cm$^{-1}$.

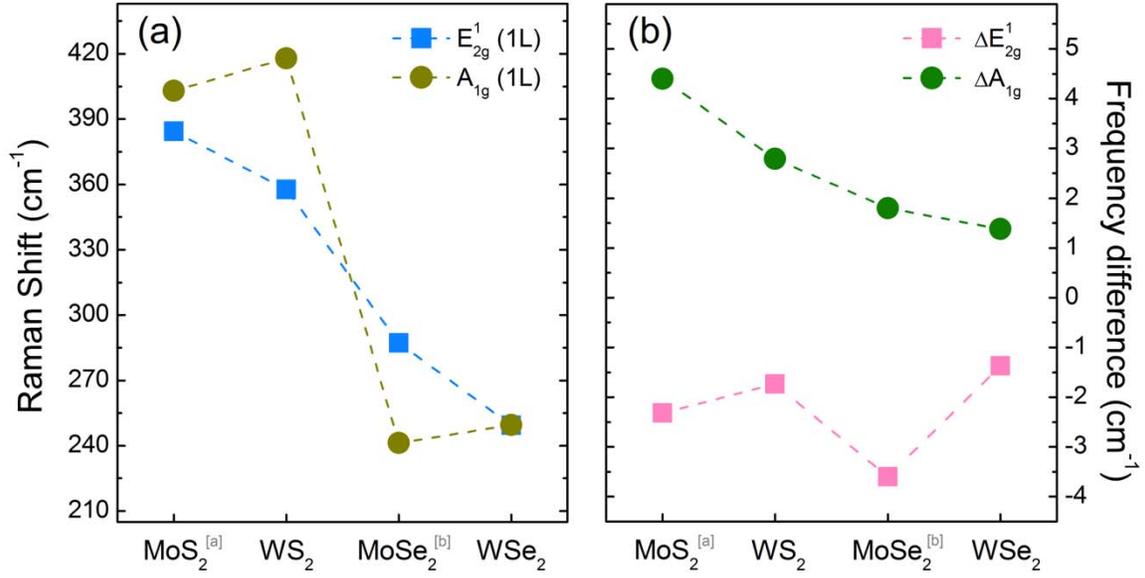

Figure 4 - (a) Frequency of the $A_{1g}$ and $E_{2g}^1$ modes in monolayer MoS$_2$, WS$_2$, MoSe$_2$ and WSe$_2$ flakes. (b) Frequency difference ($\Delta A_{1g}$ and $\Delta E_{2g}^1$) between monolayer and bulk flakes for $A_{1g}$ and $E_{2g}^1$ modes for the four compounds. We define $\Delta A_{1g}$ and $\Delta E_{2g}^1$ to be $v(A_{1g}$ (bulk)) - $v(A_{1g}$ (1L)) and $v(E_{2g}^1$ (bulk)) - $v(E_{2g}^1$ (1L)), respectively. The data for MoS$_2$ and MoSe$_2$ are obtained from Ref [42] and Ref [18], respectively.

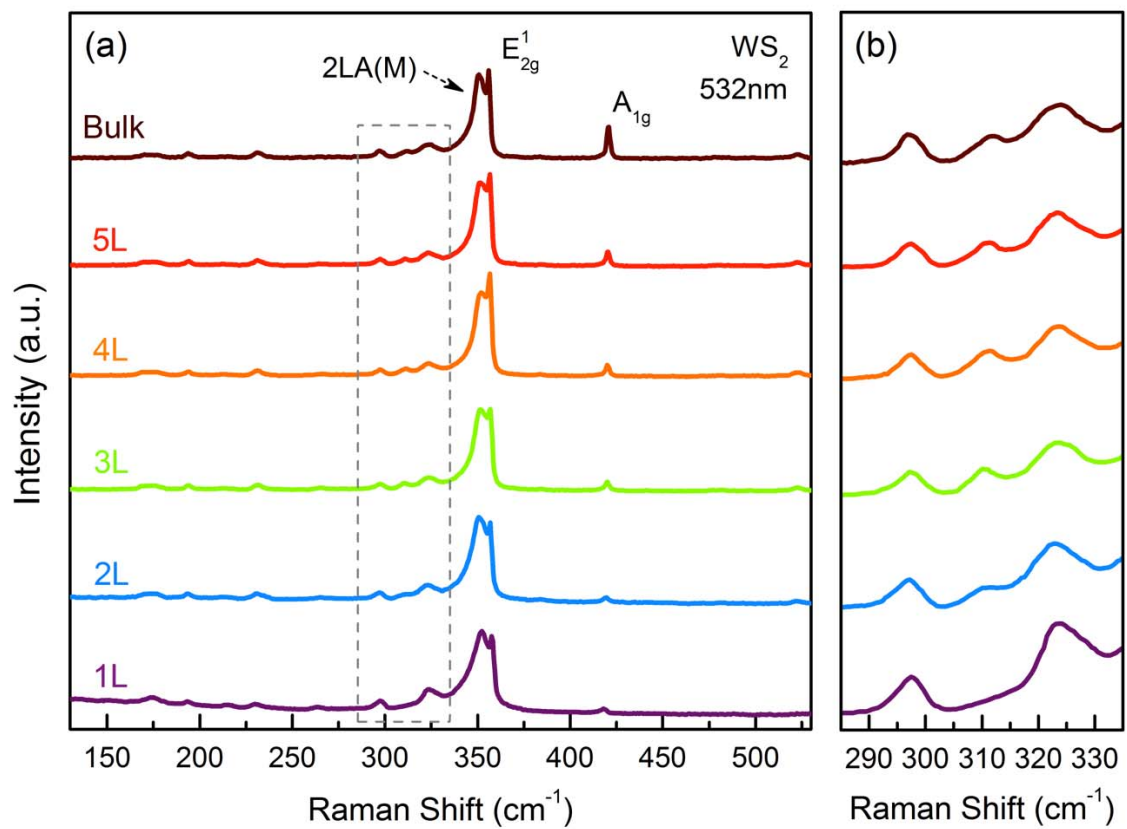

Figure 5 - (a) Unpolarized Raman spectra of 1 to 5L and bulk $WS_2$ flakes obtained with 532nm excitation. The spectra are normalized to the 2LA(M) peak and vertically offset for clarity. Part of the spectra indicated by a gray dashed rectangle in (a) is shown in larger scale in (b).

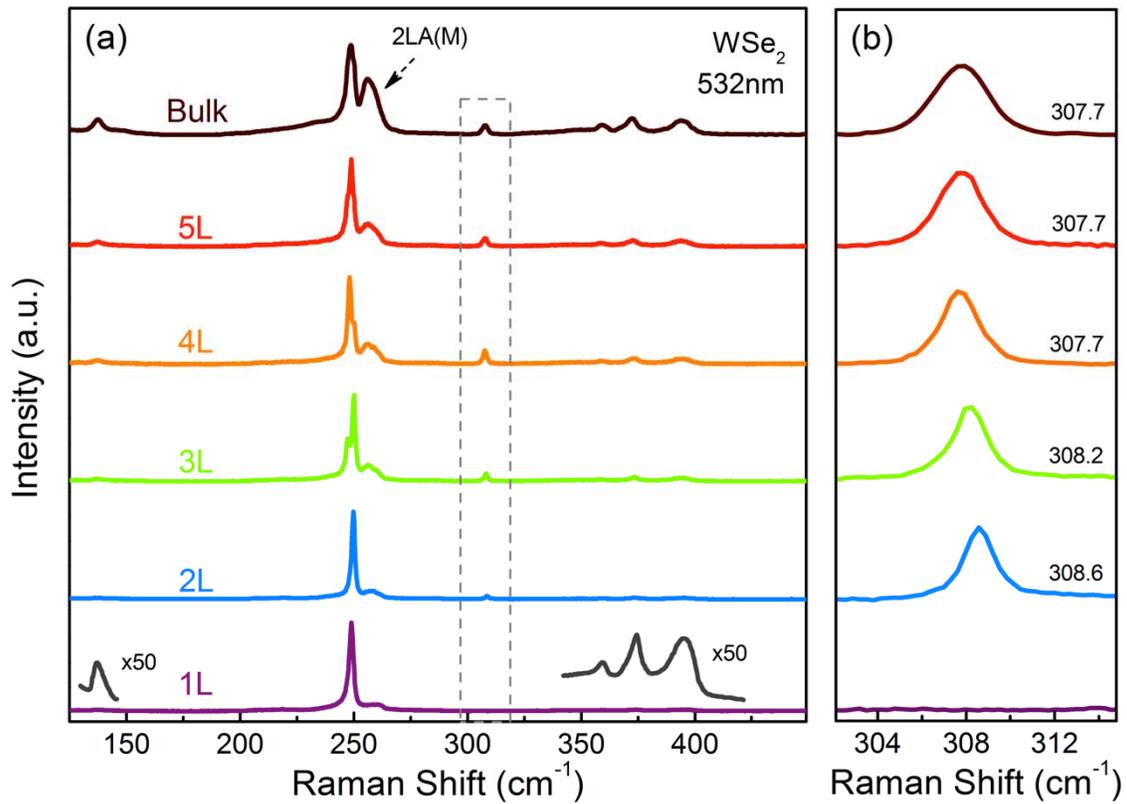

Figure 6 - (a) Unpolarized Raman spectra of 1 to 5L and bulk $WSe_2$ flakes obtained with 532 nm excitation. The spectra are normalized to the $A_{1g}$ peak and vertically offset for clarity. Part of the spectra indicated by a gray dashed rectangle in (a) is shown in larger scale in (b). The numbers in (b) indicate the peak center positions.

Table of Contents Figure

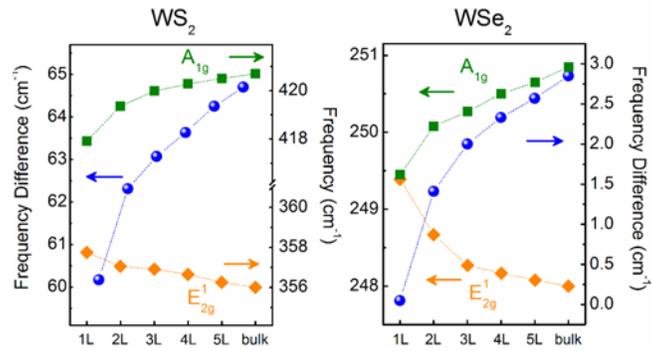

Characteristic $A_{1g}$ and $E_{2g}^1$ phonon modes for WS$_2$ and WSe$_2$ show distinct thickness dependence, which allows rapid identification of layer thickness.